\documentclass{elsart41}
\usepackage{graphics}
\usepackage{graphicx}
\usepackage{amssymb}

\begin{document}

\begin{frontmatter}

\title{
Superconducting Pairing Amplitude and Local Density of States in
Presence
of  Repulsive Centers
}

\author[TUL]{Grzegorz Litak \corauthref{Litak}}
\ead{g.litak@pollub.pl}
\author[UMCS]{Mariusz Krawiec}

\address[TUL]{
Department of Mechanics, Technical University
of Lublin,
Nadbystrzycka 36, PL-20-618 Lublin, Poland}

\address[UMCS]{
Institute of Physics and Nanotechnology Center,
Marie Curie-Sk\l{}odowska
University,
pl. M. Curie-Sk\l{}odowskiej 1, \\ PL-20-031 Lublin, Poland
}

\corauth[Litak]{Corresponding author. Tel: +48- 81- 5381573; Fax: +48- 81- 
5241004}

\begin{abstract}
We study the properties of superconductor in presence of a finite concentration 
of  repulsive centers. The superconductor is described by the negative $U$ 
Hubbard model  while repulsive centers are treated as randomly distributed 
impurities with repulsive interaction. Analyzing the paring potential and local 
density of states at impurity sites we find a wide range of the system 
parameters where the $\pi$ - like state could possibly be realized. Comparison 
of our results to the single repulsive center case is also given. 
\end{abstract}


\end{frontmatter}

A superconducting system is regarded as in the $\pi$-phase if there is a sign 
change of the order parameter between two subsystems. The simplest example is 
the junction made from two superconductors with the phase of the order 
parameter equal to $\pi$ \cite{pi_junction_Ka,pi_junction_Lo}. In this case 
Josephson current becomes negative in contrast to the usual $0$-phase junction. 
Another example are granular high-$T_c$ materials which can likely form network 
of microscopic $\pi$-junctions \cite{Sigrist} between small regions with 
different phases of the order parameter. In such systems the zero-energy 
Andreev bound states, zero-bias conductance peaks, paramagnetic Meisner effect 
and spontaneously generated currents take place 
\cite{pi_junction_Ka,pi_junction_Lo}. In our recent report \cite{Litak_Krawiec} 
we have considered a single impurity  with repulsive interaction embedded in a 
s-wave superconductor on two dimensional lattice showing conditions which 
should be fulfilled to obtain the '$\pi$' state. For a uniform embedding 
system such a state has been found in the limit of very strong repulsion at the 
impurity site. In the present paper we discuss  another possibility. Our 
superconductor possesses a finite concentration of repulsive centers with random 
distribution. With such assumptions '$\pi$' sates could appear for moderate 
values of repulsive interactions at impurities due to additional disorder in 
paring potential $\Delta_i$. 

The system is described by random $U$ Hubbard model \cite{Litak} with the 
Hamiltonian

\begin{eqnarray}
 H = \sum_{ij\sigma} \left(t_{ij} - \mu \delta_{ij}\right)
     c^+_{i\sigma} c_{j\sigma} +
     \frac{1}{2} \sum_{i\sigma} U_i n_{i\sigma} n_{i-\sigma},
 \label{Hamiltonian}
\end{eqnarray}
where $i$, $j$ label sites on a square lattice, $t_{ij} = -t$ is the hopping 
integral between nearest neighbor sites and $\mu$ is the chemical potential. 
Disorder is introduced through the random site interaction $U_i$. Depending on 
site $i$ ($i=A$ or $B$ with binary alloy distribution A$_{1-c}$B$_c$) $U_i$ 
describes attraction ($U_A < 0$) or repulsion ($U_{B} > 0)$ between electrons 
with opposite spins occupying the same site. 
\begin{figure}[htb]
\begin{center}
\includegraphics[angle=-90,width=16.5pc]{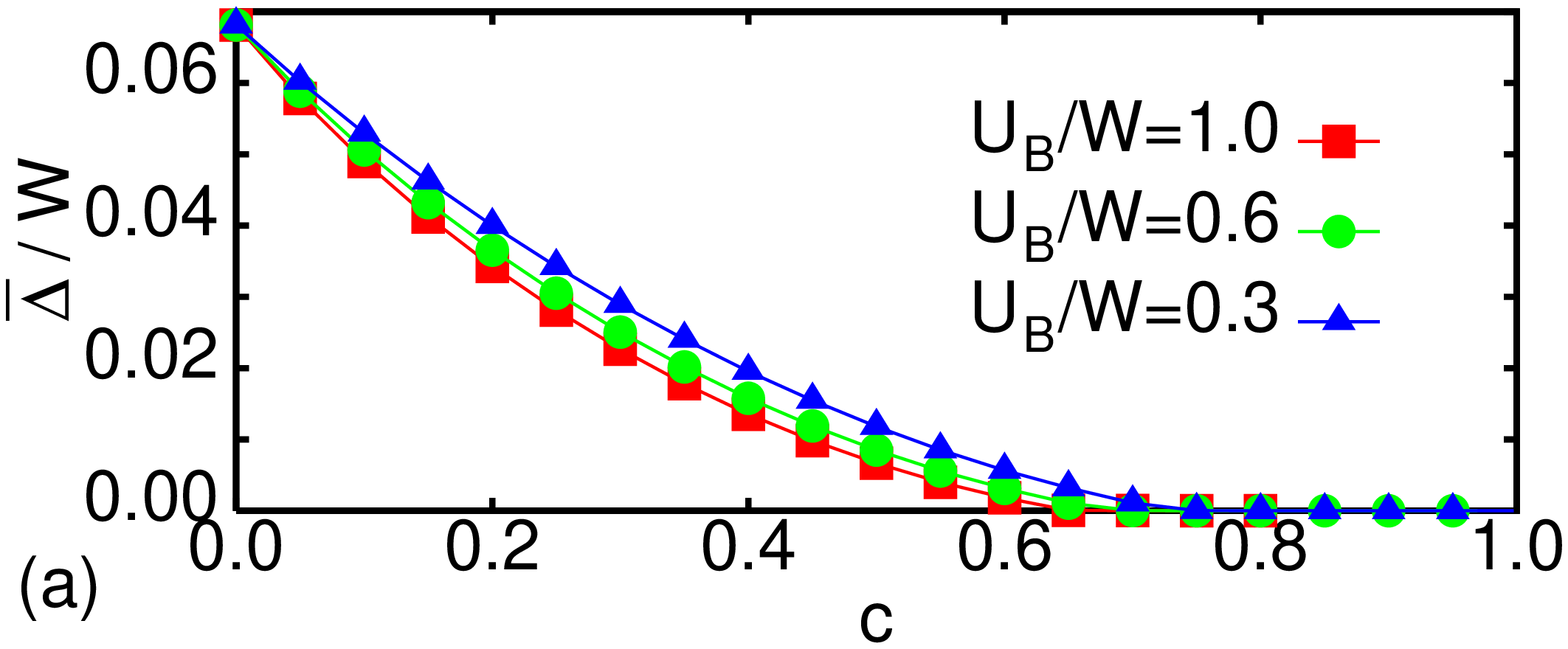}
\vspace{-0.5cm}

\includegraphics[angle=-90,width=16.5pc]{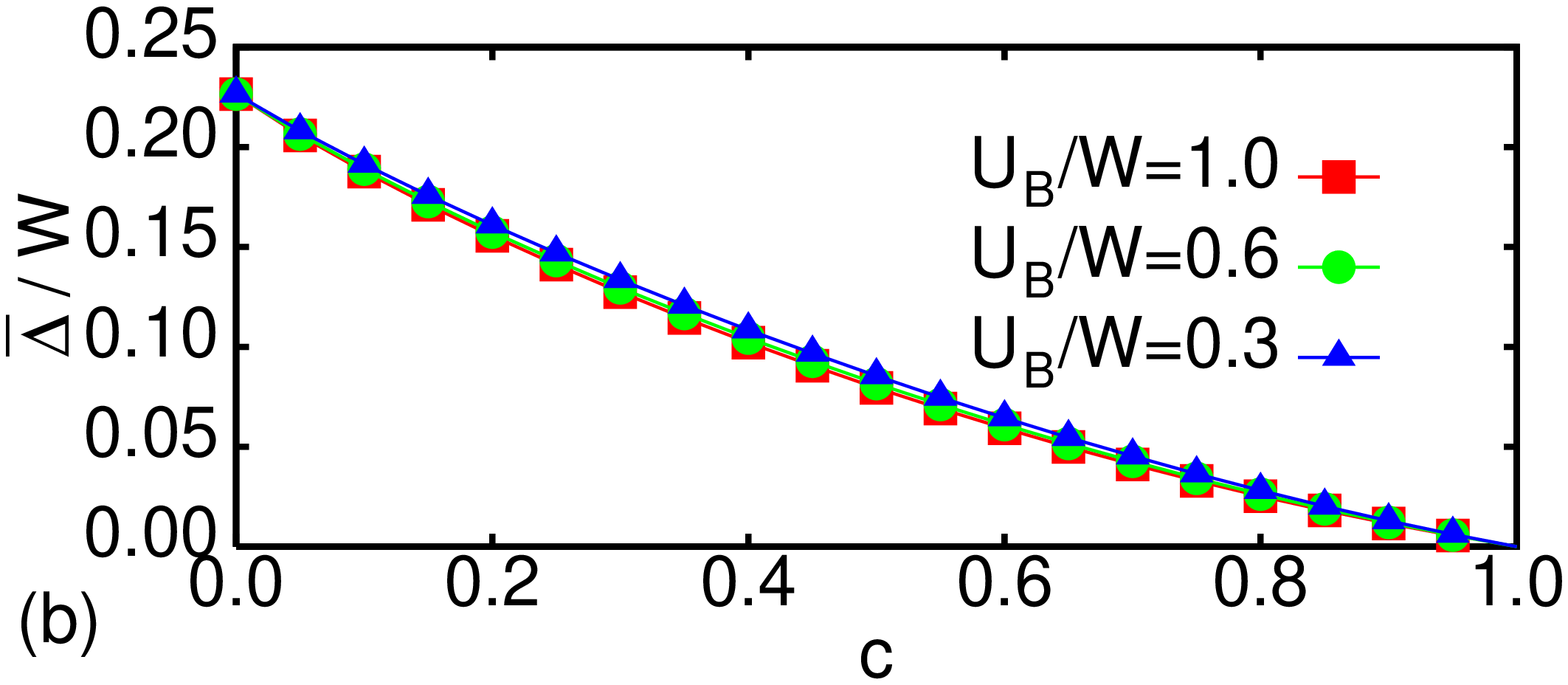}
\vspace{-0.5cm}

\includegraphics[angle=-90,width=16.5pc]{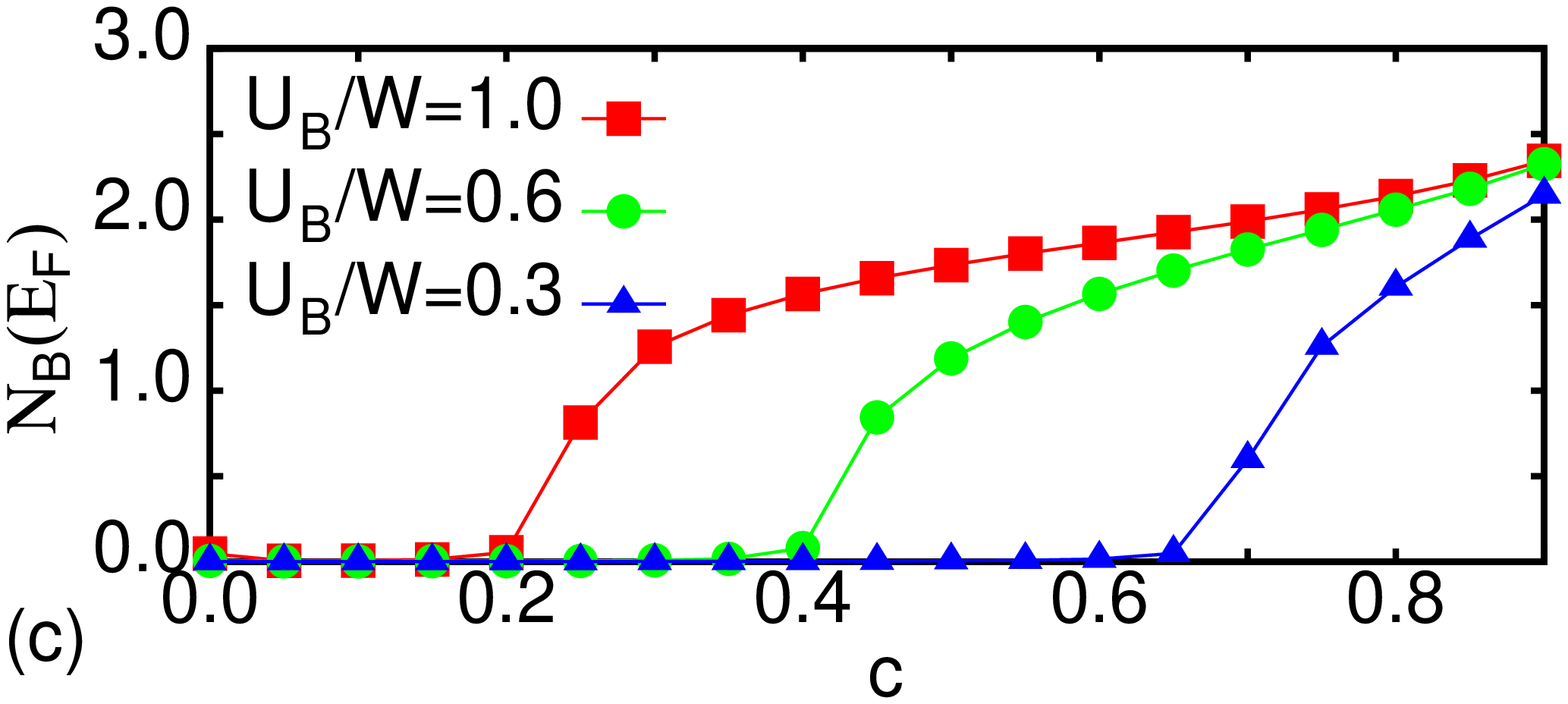}
\end{center}
\vspace{-0.5cm}

\caption{ \label{fig:1}
Superconducting order parameter $\overline{\Delta}$ versus concentration
of repulsion
impurities $c$ for $U_A/W=0.3$ (a) and  $U_A/W=0.3$ (b).
Local density of states $N_B(E_F)$ at the Fermi energy (c)
for $U_A/W=0.6$ and $U_B/W=0.3$,
0.6, 1.0.
}
\end{figure}
In the Hartree-Fock approximation \cite{Micnas} (we dropped the Hartree terms 
$U_i \langle n_{i\sigma} \rangle$, and considered only the half field system $n=1$, 
for simplicity) the corresponding Gorkov equation at zero temperature ($T=0$) 
has the following form
\begin{eqnarray}
 \sum_{j'} \left(
 \begin{array}{c}
  (\omega + \mu)\delta_{ij'} - t_{ij'} ~~~ \Delta_i \delta_{ij'} \\
  \Delta^{\ast}_i \delta_{ij'} ~~~ (\omega + \mu)\delta_{ij'} + t_{ij'}
 \end{array}
 \right)
 \hat G(j',j;\omega) = \delta_{ij},
 \label{Gorkov}
\end{eqnarray}
where
\begin{equation}
 \Delta_i \equiv U_i \chi_i =
            -  U_i \frac{1}{\pi}\int^{E_f}_{-\infty} d\omega \;
            {\rm Im} G^{12}(i,i;\omega+{\rm i} 0),
 \label{Delta}
\end{equation}
and $E_F=\mu(T=0)$ is the Fermi energy.

Using the Coherent Potential Approximation (CPA) to treat disorder  in the 
paring potential $\Delta_i$  \cite{Litak} we have found a wide region of system 
parameters 
$c$, $U_A$, $U_B$ where '$\pi$' state could exist. In that cases $\Delta_A$ and 
$\Delta_B$ were of different sign and $N_B(E_F) \neq 0$. In Fig. 1a and b we 
present the superconducting order parameter 
$\overline{\Delta}=(1-c)\Delta_A+c\Delta_B$ for attractive interaction  
$U_A=0.3$   (Fig. 1a) and $U_B=0.6$ and a number of $U_B$ repulsions. We have 
checked that in the first case  $U_A=0.3$ we have clear gap and $N_B(E_F)=0$ up 
to some critical concentration of $c=c_0$, above which the system becomes 
normal  $\overline{\Delta}=0$. In the second case  $U_A=0.6$ the systems stays 
superconducting for any $c$ but transits to '$\pi$' state for a critical value 
of concentration $c=c_1$ dependent on $U_B$. Larger $U_B$ leads to smaller 
$c_1$. Above this concentration $N_B(E_F) \neq 0$ (Fig. 1c).

For better clarity in Fig. 2a,b we have plotted local density of sates for 
$U_B=0.6$ and $U_A=0.3$ (Fig. 2a) and $U_A=0.6$ (Fig. 2b) 
\begin{figure}[htb]
\begin{center}
\includegraphics[angle=-90,width=16.5pc]{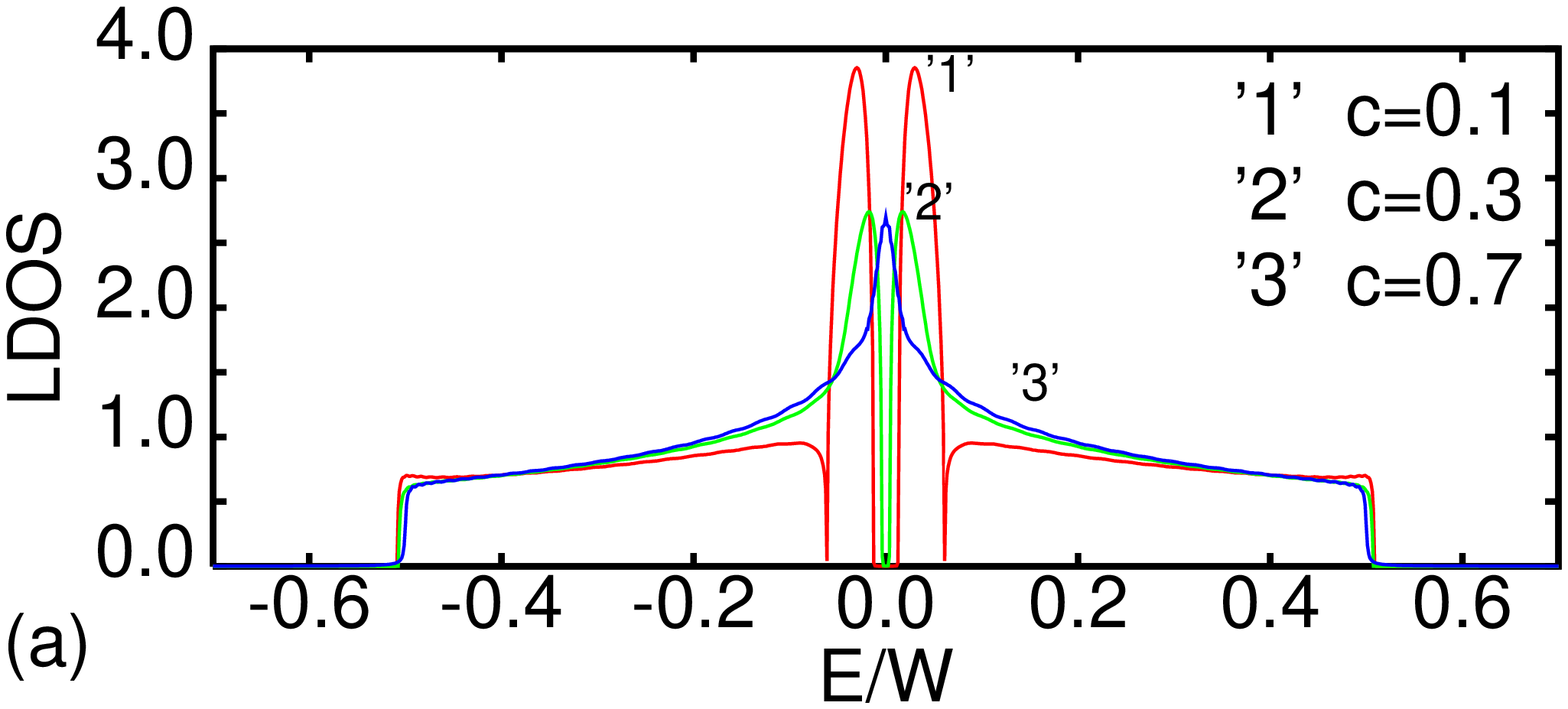}
\vspace{-0.5cm}

\includegraphics[angle=-90,width=16.5pc]{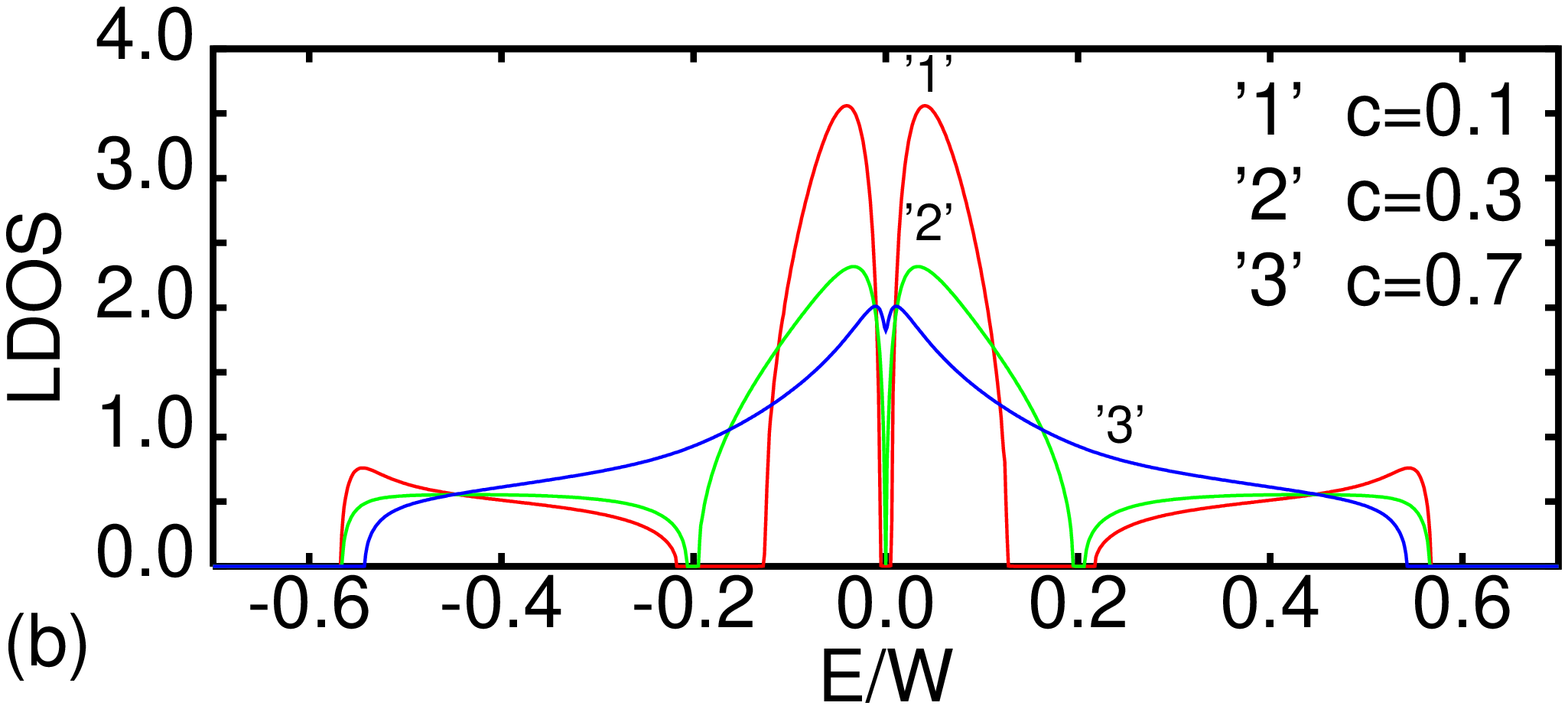}
\end{center}
\vspace{-0.5cm}

\caption{ \label{fig:2}
Local density of states (LDOS) at the B site for $U_B/W=0.6$ and $U_A/W=0.3$ (a) or
$U_A/W=0.6$ (b).} 
\end{figure}
Note that in both considered cases 
($U_A/W=0.3$ or 0.6) the impurity bands are symmetrically located with respect 
to the Fermi energy as it is in case of a single repulsive impurity 
\cite{Litak_Krawiec}. 

Our  results in Fig. 2a show transition from superconducting system $c=0.1$ and 
$c=0.3$ with a clear gap to normal state $c=0.7$. This confirms the 
results with a single impurity \cite{Litak_Krawiec} where it has been shown that 
for a relatively small bulk interaction  the '$\pi$' state is absent. On the 
other hand 
Fig. 2b shows the evolution of system with a clear gap around $E_F$ (up to 
$c=0.3$) where  $N_B(E)\neq 0$ for any $E$ but not $E=E_F$ ($N_B(E_F)=0$). 
Finally for $c=0.7$  $N_B(E_F) \neq 0$ which could possibly correspond to a 
state with $\pi$-like properties. Account for relatively large values of the 
random potential $U_i$ make Hartee-Fock approximation questionable so the 
results obtained here should be treated as qualitative. The presence of Hartree 
terms $U_i \langle n_{i\sigma} \rangle$ could also change the results in some 
range of impurity concentration $c$ forcing the transition to normal state for a 
large enough interaction $U_A$ \cite{Litak}. 

In summary we have studied properties of the superconductor with repulsive
centers and showed that for a large concentration of those centers and a large
on-center Coulomb interaction system could possibly evolve to the $\pi$ state. Two
necessary conditions for existence of the $\pi$ state, namely, the sign change
of the pairing amplitude and non-zero the density of states at the Fermi
level are fulfilled. 
 
\noindent {\bf Acknowledgements} \\
GL would like to thank Max Planck Institute for the Physics of Complex Systems 
in Dresden for hospitality.


\begin{thebibliography}{00}
\bibitem{pi_junction_Ka} S. Kashiwaya {\it et al.}, {\it Rep. Prog.
Phys.} {\bf 63} (2000) 1641.
\bibitem{pi_junction_Lo} T. L\"{o}fwander {\it et al.}, {\it Superconduct.
Sci. Technol.}  {\bf 14}, (2001) R53.
\bibitem{Sigrist} M. Sigrist and T. M. Rice, {\it Rev. Mod. Phys.} {\bf
67} (1995) 503.
\bibitem{Litak_Krawiec} G. Litak and M. Krawiec, {\it Phys. Stat. Sol.} B
{\bf 242}
(2005) 438.
\bibitem{Litak} G. Litak, B.L. Gy\"{o}rffy,  
{\it Phys. 
Rev.} B {\bf 62} (2000) 6629.
\bibitem{Micnas} R. Micnas {\it et al.}, {\it Rev. Mod. Phys.} {\bf 62}
(1991) 113.
\end{thebibliography}
\end{document}